\documentclass[pdflatex,sn-standardnature]{sn-jnl}
\pdfoutput=1
\usepackage{amsmath}
\usepackage[T1]{fontenc}
\usepackage{lmodern}
\begin{document}

\title[Dispersion Managed AFC]{Self-dispersion-managed adiabatic frequency conversion}

\author*[1]{\fnm{Dylan} \sur{Heberle}}
\email{dah378@cornell.edu}

\author[1]{\fnm{Noah} \sur{Flemens}}

\author[1]{\fnm{Connor} \sur{Davis}}

\author[2]{\fnm{Philippe} \sur{Lassonde}}

\author[2,3]{\fnm{Adrien} \sur{Leblanc}}

\author[2]{\fnm{Fran\c cois} \sur{L\'egar\'e}}

\author*[1]{\fnm{Jeffrey} \sur{Moses}}
\email{moses@cornell.edu}

\affil[1]{\orgdiv{School of Applied and Engineering Physics}, \orgname{Cornell University}, \orgaddress{\street{142 Sciences Drive}, \city{Ithaca}, \postcode{14853}, \state{New York}, \country{United States}}}

\affil[2]{INRS-emt, 1650 blvd Lionel-Boulet, Varennes, QC, J3X 1S2, Canada}

\affil[3]{Laboratoire d’Optique Appliqu\'ee, ENSTA Paris, CNRS, Ecole polytechnique, Institut Polytechnique de Paris, 91762 Palaiseau, France}

\abstract{Dispersion management of few-cycle pulses is crucial for ultrafast optics and photonics. Often, nontrivial dispersion is compensated using complex optical systems or minimized through careful design of waveguides. Here, we present dispersion-managed adiabatic frequency conversion enabling efficient downconversion of an 11.1-fs near-IR pulse to an 11.6-fs mid-IR pulse spanning an octave of bandwidth from 2--4 $\mu$m. The adiabatic frequency converter is designed to impart a constant group delay over the entire bandwidth, eliminating the need for complex dispersion management and opening a new avenue for dispersion engineering in ultrafast optics and photonics. Notably, dispersion engineering through the position-dependent conversion position of an adiabatic conversion device constitutes an additional mechanism for dispersion control in ultrafast optics and photonics while maintaining high conversion efficiency for broadband pulses.}

\maketitle

\section{Introduction}

Generating octave-spanning coherent bandwidths and compressing them to their transform-limits are among the most challenging aspects of optical systems used for ultrafast spectroscopy and control of matter, especially at the single-cycle limit. Typically, octave-spanning bandwidths are generated using super-continuum generation, noncollinear or frequency domain optical parametric amplification \cite{Chen:19, Schmidt2014May}, coherent beam synthesis \cite{manzoni2015}, nonlinear multi-mode mixing in gas-filled hollow core fibers \cite{Piccoli2021Dec}, or adiabatic frequency conversion \cite{suchowski2014adiabatic, Margules_2021}. However, as absolute bandwidths grow large, precise dispersion management in these systems plays an increasingly important role; transmission through even a few millimeters of an optical device can become a limiting technical factor. Standard compressors using bulk material, gratings, and prisms can offset large amounts of low-order dispersion but struggle to mitigate higher-order dispersion and can even exacerbate it.

To address the challenge of higher-order dispersion, modern octave-spanning systems incorporate multiple double-chirped mirror pairs or programmable dispersive filters \cite{manzoni2015} in combination with the standard approaches outlined above. However, increasingly complex dispersion management schemes adversely impact the cost, stability, flexibility, and construction time of ultrafast laser architectures. A greatly simplified single-cycle mid-IR source of CEP-stable transients could significantly extend the boundaries of ultrafast science for applications including electronics at optical clock rates \cite{Langer2018May}, attosecond science \cite{Corkum2007Jun}, and field-resolved detection of plasmonics \cite{Fischer:21}. 

Additionally, standard experiments in 2DIR spectroscopy require interferometrically stable identical pulse pairs, and more advanced experiments require the ability to impress differences on these pairs such as relative phase, amplitude, and spectral content. An interferometer produces identical mid-IR pulse pairs but lacks stability and introduces timing jitter. A spatial light modulator inside a grating pair or an acousto-optic programmable dispersive filter (AOPDF) would enable arbitrary shaping of post-generation pulses; however, such devices are usually limited to sub-octave-spanning bandwidths. Piecewise dispersion compensation is possible through coherent pulse synthesis schemes but requires substantial engineering \cite{manzoni2015}.

Several researchers have incorporated dispersion management and pulse shaping into nonlinear frequency conversion via chirped quasi-phase-matched (QPM) periodically poled gratings. Arbore et al. \cite{ArboreTheory:97} proposed compressing optical pulses using second-harmonic generation (SHG) with a chirped QPM grating and demonstrated pulse compression of 110-fs FWHM frequency-doubled Ti:Sapphire pulses \cite{ArboreExp:97}. Furthermore, Imeshev et al. demonstrated pulse shaping and arbitrary dispersion management in SHG \cite{Imeshev:98, Imeshev:00, ImeshevReview:00} and extended the platform to include difference frequency generation (DFG) in the undepleted pump and unamplified signal regimes \cite{Imeshev:01}. Charbonneau-Lafort et al. \cite{Charbonneau-Lefort:05, Charbonneau-Lefort:08} used tandem chirped QPM optical parametric amplification (OPA) devices to engineer both the gain and group delay.

\begin{figure}
    \centering		    
        \includegraphics[width=4.6 in]{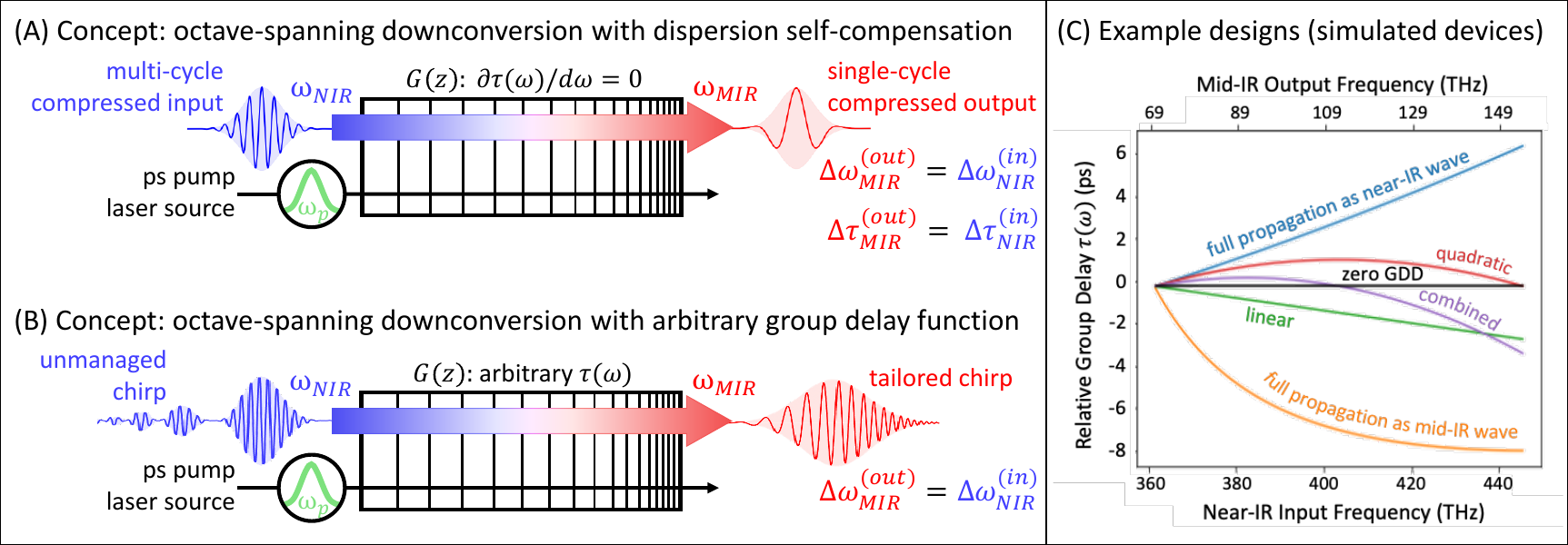}
        \caption{Self-dispersion managed ADFG concept: a compressed multi-cycle near-IR wave undergoes bandwidth and phase-conserving adiabatic down-conversion to a compressed, CEP-stable, single-cycle mid-IR pulse in a chirped QPM grating with a local poling frequency G(z) designed to impart zero GDD and higher order dispersion.}
    \label{fig:Concept}
\end{figure}

While these studies have simplified ultrafast sources, the use cases are limited by fundamental and signal sources (in the case of SHG and OPA respectively) and linearity (in the case of DFG). We extend dispersion management and pulse shaping to adiabatic frequency conversion \cite{Suchowski2011} which enables nearly complete photon-number transfer between spectral regions while preserving the absolute bandwidth of the pulse and can be configured for adiabatic difference frequency generation (ADFG) or adiabatic sum frequency generation (ASFG) greatly increasing the output spectral range. Here, we demonstrate a highly efficient adiabatic frequency downconverter that intrinsically manages its own large material dispersion. The device concept is outlined in Fig. \ref{fig:Concept}a. The ADFG device converts a multi-cycle, compressed near-IR pulse at Ti:Sapphire wavelengths (680-820) to a single-cycle, compressed mid-IR pulse (2–4 µm). This device allows a linear transfer of amplitude and phase from the near-IR pulse to an octave-spanning bandwidth in the mid-IR without the need for additional complex dispersion management. Furthermore, our concept can accommodate arbitrary dispersion functions as shown in Fig. \ref{fig:Concept}b where a near-IR pulse with high-order dispersion is converted to chirped mid-IR pulse with dispersion that can be easily compressed with a typical compressor. The group delays of example designs are plotted in \ref{fig:Concept}c including constant group delay (our device), linear group delay (constant group delay dispersion (GDD)), quadratic group delay (constant third-order dispersion (TOD)), and combined linear and quadratic group delay (constant GDD and TOD). Possible group delay functions are bounded by the group delay curves representing complete propagation as a near-IR pulse (blue curve) and as a mid-IR pulse (orange curve). We note our technique is not restricted to adiabatic difference frequency generation and can be extended to adiabatic sum frequency generation and adiabatic four-wave mixing.

Below, we report a 2-cm LiNbO\textsubscript{3} device that converts a transform-limited pulse in the near-IR to an octave-spanning transform-limited pulse in the mid-IR. The result is a compressed 11.1-fs near-IR pulse converted to an 11.6-fs mid-IR pulse with high photon conversion efficiency in the presence of a strong, narrowband pump. This constitutes a nearly complete elimination of GDD and a dramatic improvement over the $>10^4$ fs$^2$ GDD imparted by a standard, quasi-linear QPM domain poling function, as used in \cite{Krogen2017Apr}. 

\section{ADFG concept}

\subsection{ADFG}

Ultrafast ADFG is a conversion technique for efficient down-conversion of photons over a wide bandwidth \cite{suchowski2014adiabatic, Margules_2021}. It is achieved using a monotonic, slowly varying position-dependent poling frequency $G(z)=2\pi/\Lambda(z)$ to create a chirped QPM device where $\Lambda(z)$ is the domain period of the structure at the propagation coordinate $z$. Each input angular frequency $\omega$ is locally converted within the device at a position $z_c(\omega)$ when the phase matching condition $\Delta k(\omega)=\Delta k_0(\omega) - G(z_c(\omega))=0$ where $\Delta k_0(\omega)$ is the bulk wavevector mismatch of the DFG process. Here, $\Delta k_0(\omega) = k_{NIR}(\omega) - k_{MIR}(\omega) - k_p$ where $k_{NIR}(\omega)$, $k_{MIR}(\omega)$, and $k_p$ are the near-IR, mid-IR, and pump wavevectors. Note that since the pump is very narrowband, there is a one-to-one correspondence between near-IR and mid-IR frequencies. When the poling function varies slowly enough, the adiabatic condition is achieved and photons at each frequency undergo unidirectional conversion to the difference frequency field, in this case the mid-IR field. The efficiency of this process approaches saturation at unity with increasing pump power \cite{Moses:12}.

ADFG has been used to produce single-cycle, energetic pulses when accompanied by a complex dispersion management scheme incorporating grism pairs and a programmable dispersive filter \cite{Krogen2017Apr}. That work also showed the ability of ADFG to preserve complex pulse features through the conversion process. However, the device required a complicated phase mask applied by a programmable dispersive filter to compress the pulses. This was required because of the complicated dispersion profile imparted by the conversion device due to the longitudinally varying conversion positions in addition to the bulk material dispersion.

\begin{figure}
    \centering		    
        \includegraphics[width=3.4 in]{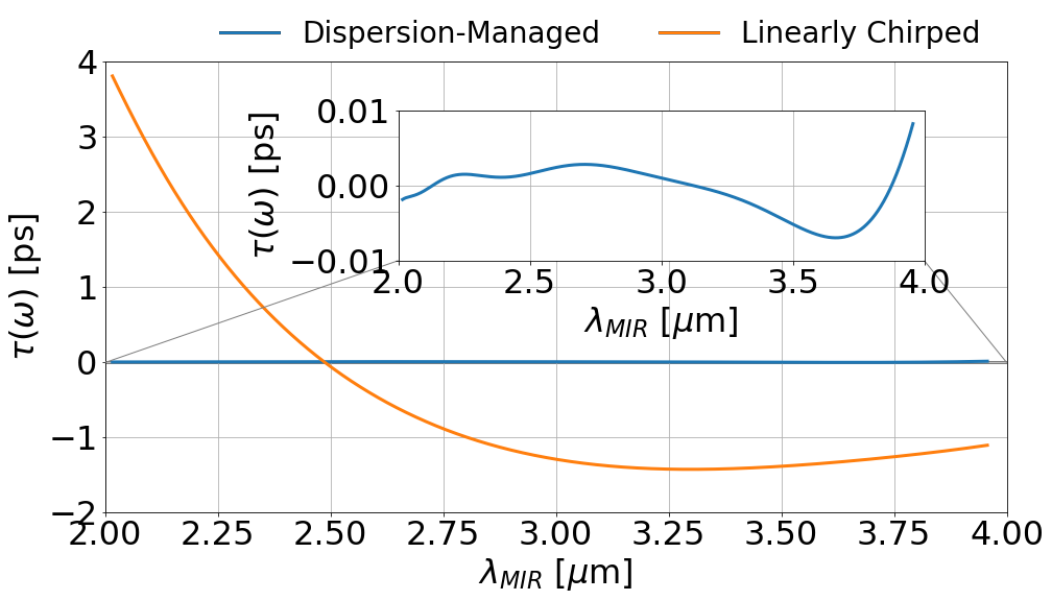}
        \caption{Sub-20-fs peak-to-peak residual group delay for the dispersion managed device compared to >5 ps for a device of identical conversion bandwidth and length, but with a standard, linearly chirped G(z).}
    \label{fig:DMSimulationPhase}
\end{figure}

The dispersion that results from a monotonically varying aperiodic grating has been studied for both adiabatic optical parametric amplification (AOPA) \cite{Charbonneau-Lefort:05} and ADFG \cite{Krogen2017Apr} devices. The group delay $\tau(\omega)$ as a function of conversion position $z_c(\omega)$ is given by:
\begin{equation}
\tau(\omega)=k_{NIR}^{'}(\omega) z_c (\omega)+k_{MIR}^{'}(\omega)(L-z_c(\omega))
\label{eq:gd}
\end{equation}
\noindent where the primes denote differentiation with respect to $\omega$. 
Intuitively, this equation states that the total group delay is the sum of that acquired by the input near-IR frequency up to the conversion position plus the group delay acquired by mid-IR frequency after the conversion position. A consequence of Eq. \ref{eq:gd} is that the device dispersion is highly dependent on the design. A linear poling function $G(z)$ is the simplest design to maintain the adiabatic condition over a wide bandwidth, and yet it can result in a non-trivial group delay that can not be compensated without the aid of a pulse shaper. Fig. \ref{fig:DMSimulationPhase} shows that for a linearly chirped 2-cm LiNbO\textsubscript{3} device, the resulting group delay (orange curve) spans several picoseconds with large amounts of high order dispersion indicated by the line shape.

\subsection{Dispersion Engineering}

The frequency-dependent and localized conversion of ADFG can be used to cancel the device's own material dispersion while generating an octave-spanning bandwidth in the mid-IR. A similar concept was described theoretically for adiabatic OPA where two devices were used in tandem to engineer a dispersion profile for the seeded field \cite{Charbonneau-Lefort:05}. Since the desired output of and ADFG device is the unseeded field, this can be accomplished in a single device. To cancel the material dispersion of 2 cm of LiNbO\textsubscript{3}, we design a QPM grating such that the group delay profile is flat, i.e., $d\tau(\omega)/d\omega=0$, to minimize group delay dispersion and higher order dispersion imparted by the device (Fig. \ref{fig:DMSimulationPhase}, blue). 

To design the device, we solve Eq. \ref{eq:gd} for $z_c(\omega)$ given $\tau(\omega)=C$ where $C$ is a constant. The constant can take any value that lies between $k_{NIR}^{'}(\omega) L$ and $k_{MIR}^{'}(\omega) L$ for all relevant values of $\omega$ such that $z_c(\omega)$ is monotonic. The grating can then be constructed by choosing a poling period $\Lambda(z)$ such that $G(z_c(\omega))=k_0(\omega)$ at each $z_c(\omega)$. Since $z_c(\omega)$ is continuous but the grating structure is made of discrete domains, this can be accomplished by sampling $z_c(\omega)$ according to the current position in the device after placing each domain. 

\begin{figure}
    \centering		    
        \includegraphics[width=3.4 in]{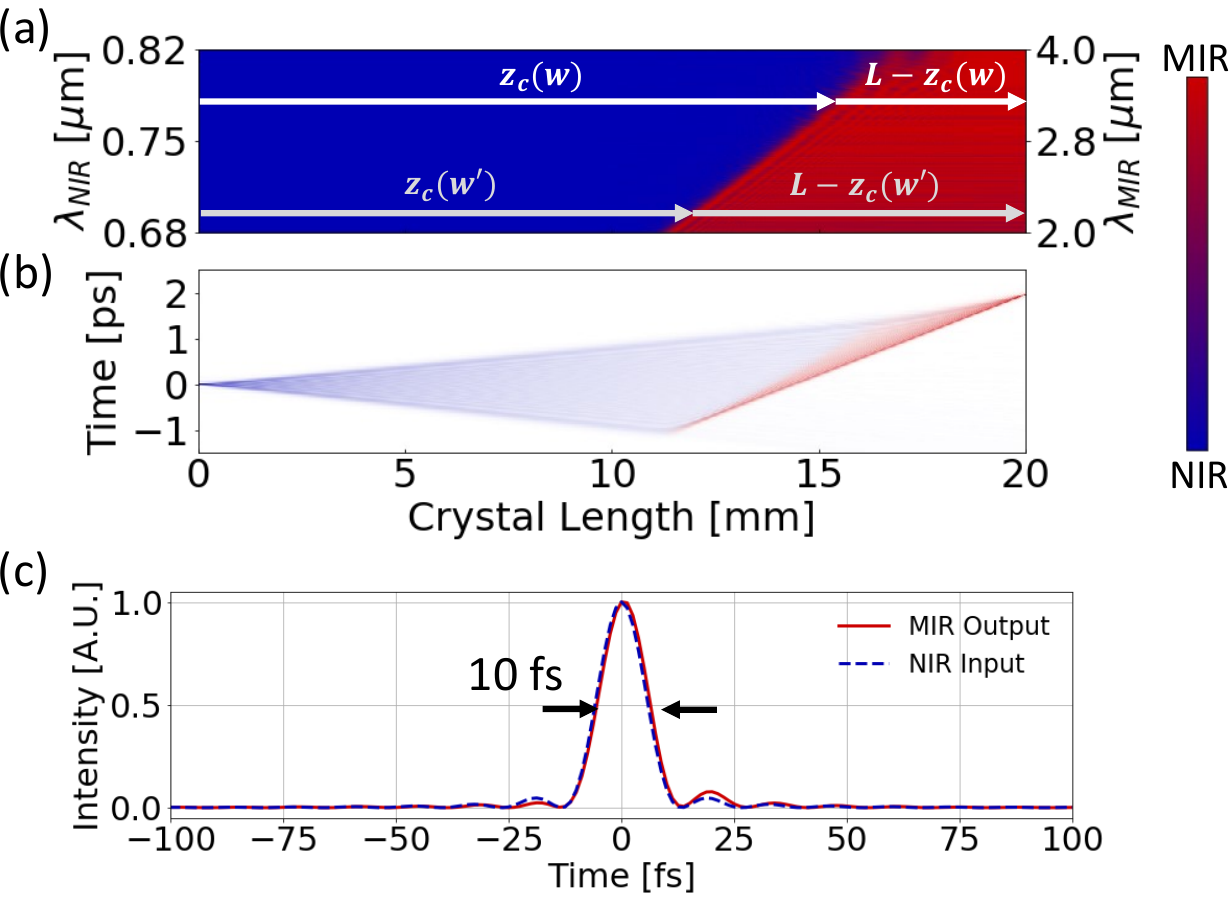}
        \caption{Propagation simulations of the device under study, showing (a) the position-dependent conversion of each spectral component of an input 10-fs near-IR field, (b) temporal evolution dynamics illustrating stretching of the input near-IR pulse followed by simultaneous conversion to the mid-IR and recompression to a 10-fs duration. (c) Mid-IR output pulse intensity (solid red) compared against initial compressed near-IR input pulse (dashed blue).}
    \label{fig:DMSimulation}
\end{figure}

Fig. \ref{fig:DMSimulation}a shows a simulation of the near-IR and mid-IR spectra as a function of propagation in our dispersion-managed device. The propagation-dependent conversion position of each mid-IR wavelength in the device can be seen from the sharp conversion transition (i.e. blue to red). In the time domain, this results in a compressed, 10-fs near-IR input converting directly to a compressed, 10-fs mid-IR output (Fig. \ref{fig:DMSimulation}b,c). Conveniently, the device acts as its own stretcher/compressor, broadening the input near-IR pulse to several picoseconds prior to conversion, keeping the peak intensity low. Subsequently, the generated mid-IR pulse self-compresses back to 10 fs  (Fig. \ref{fig:DMSimulation}). Furthermore, $\tau(\omega)$ can be engineered to suit specific applications such as compensation of external sources of dispersion so that the input pulse itself need not be compressed. 

\section{Experimental Results}

\subsection{Setup}

\begin{figure*}
    \centering		    
        \includegraphics[width=\textwidth]{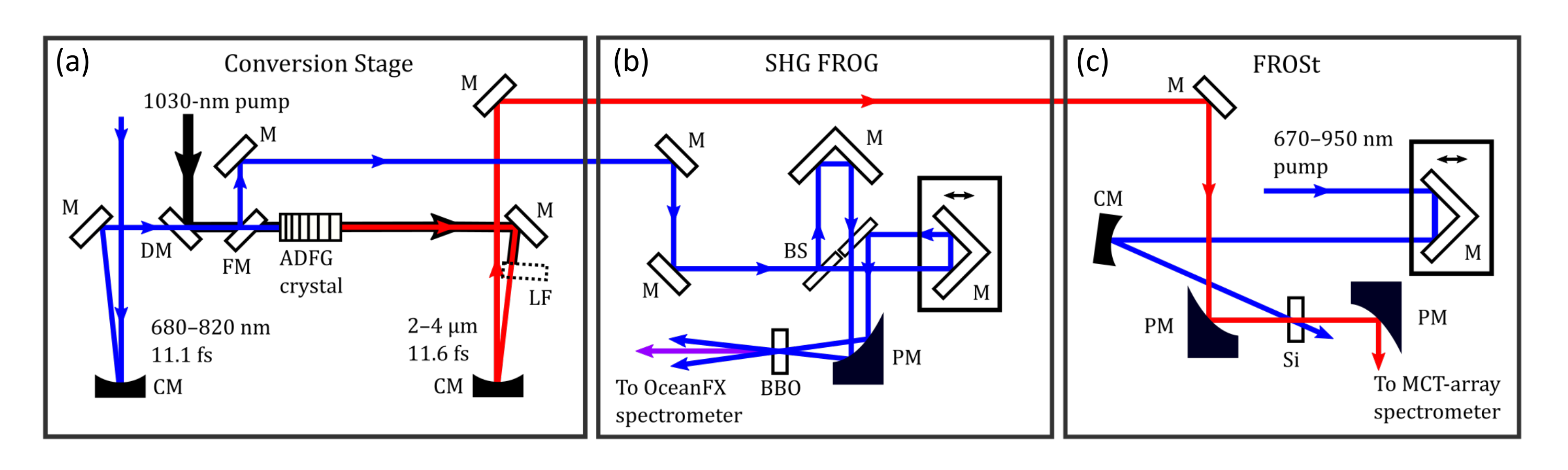}
        \caption{Experimental setup. (a) Conversion stage with the dispersion-managed ADFG device. The near-IR beam is focused into the ADFG crystal using a concave mirror (CM) and combined with the 1030-nm pump using a dichroic mirror (DM). The generated mid-IR light is collimated with a second CM. (b) SHG FROG measurement stage. The near-IR pulse is diverted before the ADFG stage using a flip mirror (FM) and measured using a dispersion-balanced SHG FROG setup. (c) FROSt measurement stage used to characterize the mid-IR pulse generated from the compressed near-IR pulse. The mid-IR pulse is separated from the residual pump and near-IR beams using a longpass filter (LF). Additional optics include mirrors (M), beamsplitter (BS), and parabolic mirrors (PM).}
    \label{fig:experimentalSetup}
\end{figure*}

The laser source consists of a 1030-nm Yb:YAG laser with two outputs: a 900-fs low-power (2 \textmu J) output and a 3-ps high-power (17 mJ) output. The 900-fs low-power output drives white-light (WL) generation in a 13-mm YAG crystal, from which a bandwidth of 670--950 nm is selected and amplified to 50 \textmu J in a series of noncollinear optical parametric chirped pulse amplifiers pumped with frequency-doubled light from the high-power output. The amplified near-IR pulse is split into several paths. The first is sent to the adiabatic frequency converter to generate the 2--4 \textmu m mid-IR pulse. A 4f fourier-domain pulse shaper (PhaseTech, Inc.) is used to spectrally filter the near-IR pulse to 680--820 nm and compress it at the input of the adiabatic frequency conversion stage. A second near-IR beam, compressed with a chirped mirror pair, is used as the pump pulse for frequency-resolved optical switching (FROSt), described below, which is used to characterize the generated mid-IR pulse.

The ADFG setup is shown in Figure \ref{fig:experimentalSetup}a. The compressed 680--820-nm near-IR pulse is combined with 200 \textmu J of 1030-nm pump light in the ADFG crystal generating a compressed 2--4-\textmu m mid-IR pulse. The near-IR beam is focused into the frequency conversion crystal with a 50-cm concave mirror and combined with the pump pulse using a shortpass dichroic mirror. A longpass filter (Andover 1650-nm longpass filter, 1.1-mm Si) is used to separate the mid-IR pulse from the residual near-IR and pump light. The mid-IR light is collimated using a 50-cm concave mirror. Through the ADFG device, we achieve a nearly flat 70\% photon conversion efficiency across the input near-IR spectrum.

The near-IR pulse is characterized at the input of the ADFG stage using SHG FROG as shown in Fig. \ref{fig:experimentalSetup}b. The SHG FROG stage is a noncollinear, dispersion-balanced autocorrelator using BBO for SHG. SHG FROG traces are constructed by measuring the second harmonic spectrum as a function of pulse delay using an OceanFX spectrometer. Using these measurements, the near-IR pulse is compressed by adding a phase with the 4f pulse shaper.  

The generated octave-spanning mid-IR pulse is characterized using FROSt. FROSt is a phase-matching-free characterization technique that was recently used to characterize two-octave-spanning infrared pulses \cite{Leblanc_2021, Longa:22}. In FROSt, a pump pulse is used to excite free-carriers in a semiconductor sample creating a sharp absorption edge in time which acts as an optical gate for characterizing a probe pulse. The probe pulse spectrum is measured as a function of pump-probe delay, allowing the electric field of the probe pulse to be reconstructed using a ptychographic algorithm \cite{Leblanc_2021}. The FROSt setup is shown in Fig. \ref{fig:experimentalSetup}c. We use a 10-\textmu J near-IR pulse compressed to 14-fs FWHM to pump the free-carrier excitation in a 1-mm Si sample. The pump and probe pulses are combined at the FROSt sample. The probe is focused to a ~44-\textmu m 1/$e^2$ beam diameter at the sample. To achieve a strong, uniform free-carrier excitation across the probe beam, the pump is focused to a ~275-\textmu m 1/$e^2$ beam diameter. The mid-IR spectrum is measured using a MCT-array spectrometer as a function of pump-probe delay.

\subsection{Results}

\begin{figure*}
    \centering		    
        \includegraphics[width=4.7 in]{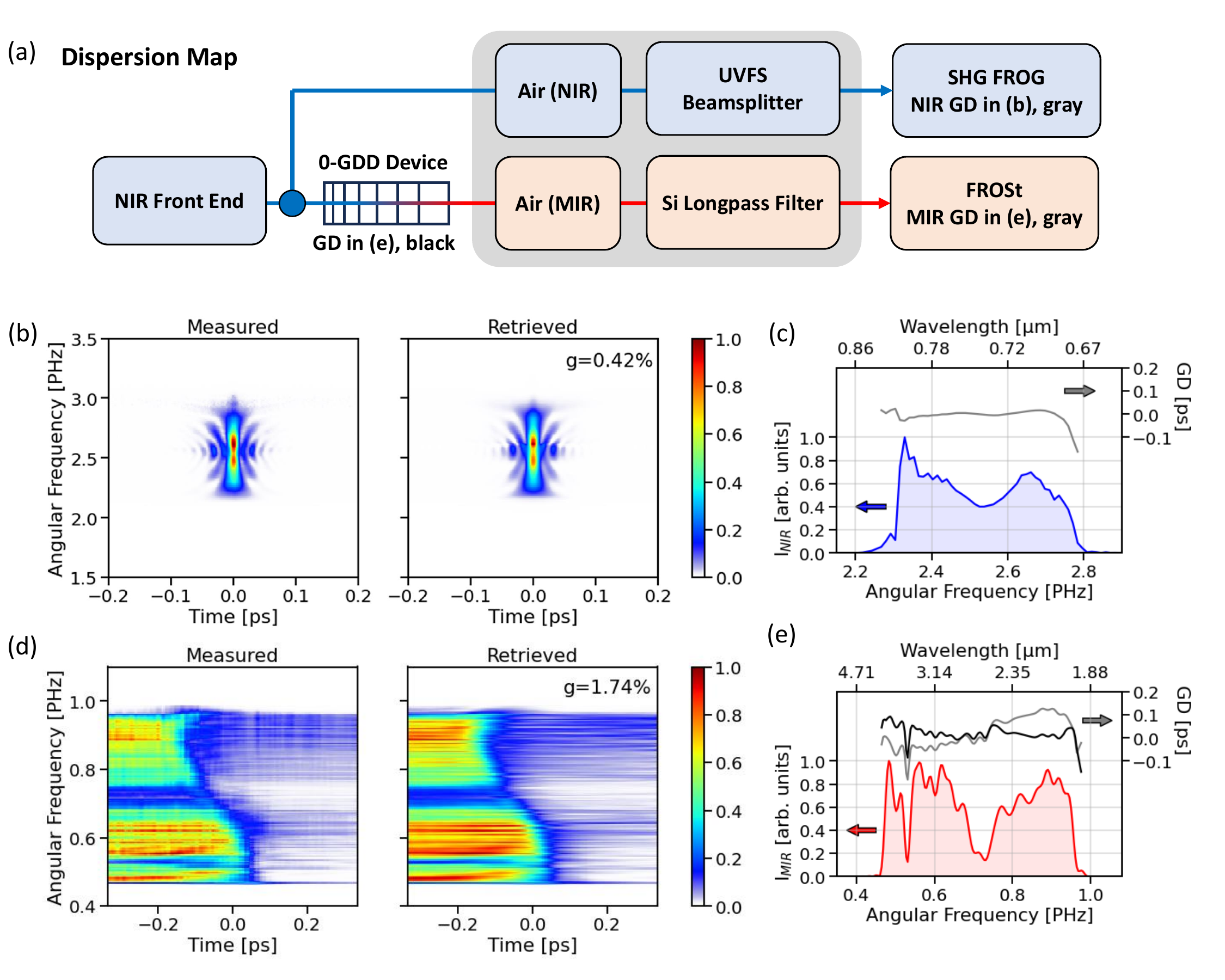}
        \caption{(a) Outline of experimental setup indicating components (shaded region) that must be considered in the calculation of the group delay imparted by our zero-GDD device using the near-IR and mid-IR pulse measurements. (b) Measured and retrieved SHG FROG traces for the near-IR pulse. (c) The retrieved near-IR spectral intensity (blue) and group delay (gray). (d) Measured and retrieved FROSt traces for the mid-IR pulse. (e) The retrieved mid-IR spectral intensity (red) and group delay (gray) with the calculated GD of our zero-GDD device (black). The GD imparted by the device is the difference between the gray GD curves in (e) and (c) accounting for the components in the shaded region shown in (a).}
    \label{fig:Results}
\end{figure*}

 The measured and retrieved SHG FROG traces for the near-IR pulse are shown in Fig. \ref{fig:Results}b with the normalized RMS error, g, described in \cite{trebino2012frequency}. Figure \ref{fig:Results}c shows the retrieved near-IR spectrum (blue) and group delay (gray). The measured and retrieved FROSt traces for the mid-IR pulse are shown in Fig. \ref{fig:Results}d. Figure \ref{fig:Results}e shows the retrieved mid-IR spectrum (red) and group delay (gray). A number of modulations are observed across the mid-IR spectrum and have been identified with a parasitic OPA process that occur at the end of the device in a region that is non-critical to the dispersion properties of the device.

The group delay imparted by the device is determined by comparing the near-IR spectral phase measured with SHG FROG and the output mid-IR spectral phase measured with FROSt while accounting for all the dispersive components in the paths to the measurement stages. These components are shown by the shaded region in \ref{fig:Results}a and include a 1-mm fused silica beamsplitter used in the SHG FROG stage, a 1.1-mm silicon (Si) longpass filter used to isolate the mid-IR pulse, and the air in the beam paths to the measurement devices. Taking these into account, the spectral phase imparted by our device $\phi_D$ and related group delay $\tau(\omega)=(d\phi_D (\omega))⁄d\omega$ is calculated by $\phi_D=\phi_{MIR}-\phi_{NIR}-\phi_{optics}-\Delta\phi_{air}$ where $\phi_{MIR} (\phi_{NIR})$ is the measured spectral phase of the mid-IR (near-IR) pulse, $\phi_{optics}$ is the measured spectral phase of the longpass filter minus the analytic spectral phase of the beamsplitter calculated from the Sellmeier equation \cite{malitson:65}, and $\Delta\phi_{air}$ is the difference in phase experienced by the mid-IR and near-IR pulses as they travel through air to their respective measurement setups. 
Note, $\phi_{NIR}$ and $\phi_{MIR}$ directly correspond to the gray group delay curves shown in Fig. \ref{fig:Results}c and \ref{fig:Results}d respectively.

\begin{figure*}
    \centering		    
        \includegraphics[width=4.7 in]{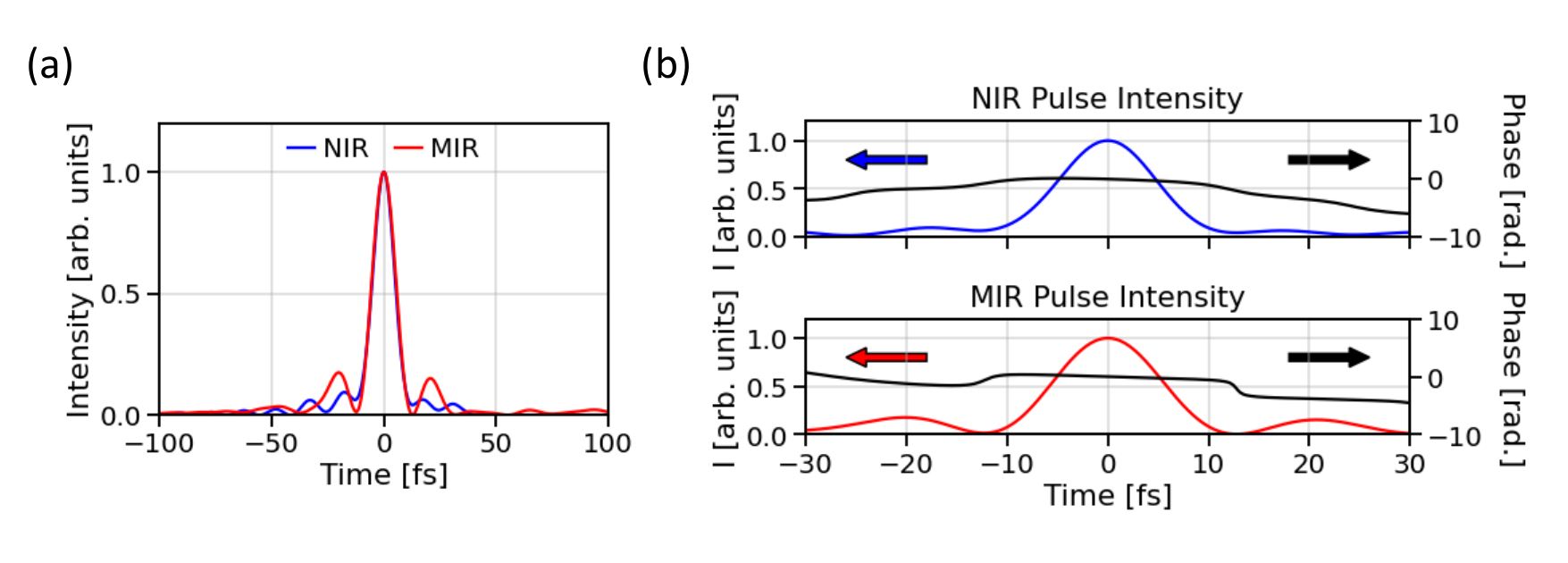}
        \caption{(a) Retrieved pulse intensities of the near-IR input (blue) and mid-IR output (red). The 11.1-fs near-IR pulse is converted to an 11.6-fs mid-IR pulse. (b) Zoomed-in plots of the near-IR (top) and mid-IR (bottom) pulse intensities with temporal phases (black).}
    \label{fig:ResultsTime}
\end{figure*}

The group delay imparted by our device is shown as the black curve in Fig. \ref{fig:Results}e. Over the target bandwidth (2–4 \textmu m), the group delay remains nearly flat, only deviating near the edges of the spectrum and at 0.7 and 0.53 PHz where dips occur in the power spectrum. We believe the dip at 0.7 PHz is caused by water absorption and the dip at 0.53 PHz is caused by parasitic effects due to the uniform poling period at the end of the ADFG grating. To demonstrate the performance of the device, Fig. \ref{fig:ResultsTime}a compares the mid-IR output pulse (red) using our compressed near-IR pulse (blue). The group delay imparted by the device results in an 11.6-fs pulse (FWHM duration), which is 1.05x the input near-IR pulse. Zoomed-in images with the retrieved phase are shown in Fig. \ref{fig:ResultsTime}b.

\section{Discussion}

In summary, we have demonstrated the proof-of-principle of a concept for generating single-cycle mid-IR pulses efficiently by adiabatic down-conversion of a few-cycle near-IR source, in a monolithic device imparting near-zero GDD. Alternatively, the device could be designed to impart a desired dispersion profile in combination with a small amount of bulk material dispersion. This would enable high energy frequency conversion by avoiding high peak intensities in the device.

We believe this work is an important step in advancing the state of the art for ultrafast lasers and photonics as well as the study of ultrafast science. By compensating for its own dispersion and/or the dispersion of other components of the laser architecture, the dispersion managed ADFG concept can greatly reduce the complexity, cost, and time investment of building ultrafast laser systems. It also serves to enhance the modularity and compactness of ultrafast frequency conversion stages by eliminating the need for the sophisticated dispersion compensation stages. Such stages introduce instability by increasing extra beam paths and optics and have to be tailored to the specific conversion device, making them inflexible. Finally, dispersion-engineered adiabatic frequency conversion provides another mechanism to tune the dispersion of integrated photonics, which is typically achieved through geometric dispersion of waveguides, greatly increasing the tunability and range of ultrafast photonic conversion devices.

\backmatter

\bigskip

\bibliography{DispersionManagedADFG}

\end{document}